\documentclass[aip,jcp,amsmath,amssymb,reprint,floats]{revtex4-1}

\usepackage{graphicx}
\usepackage{bm}
\usepackage{dcolumn}
\usepackage{bm}

\usepackage[T1]{fontenc}
\usepackage{mathptmx}
\usepackage{amssymb}
\usepackage{color}
\usepackage{amsmath}
\usepackage{enumitem}

\newcommand{\beq}{\begin{equation}}
\newcommand{\eeq}{\end{equation}}
\newcommand{\beqar}{\begin{eqnarray}}
\newcommand{\eeqar}{\end{eqnarray}}
\newcommand{\bea}{\begin{eqnarray}}
\newcommand{\eea}{\end{eqnarray}}
\newcommand{\bcen}{\begin{center}}
\newcommand{\ecen}{\end{center}}

\newcommand{\ave}[1]{\left< #1 \right>}

\newcommand{\Hop}{\hat H}

\newcommand{\aop}{\hat{a}}

\newcommand{\cop}{\hat{c}}
\newcommand{\dop}{\hat{d}}

\newcommand{\Iop}{\hat{I}}
\newcommand{\nop}{\hat{n}}
\newcommand{\Aop}{\hat{A}}

\newcommand{\iop}{\hat{\text{i}}}
\newcommand{\jop}{\hat{\text{j}}}
\newcommand{\kop}{\hat{\text{k}}}
\newcommand{\iopm}{\hat{\text{i}}_m}
\newcommand{\iopn}{\hat{\text{i}}_n}
\newcommand{\jopm}{\hat{\text{j}}_m}
\newcommand{\jopn}{\hat{\text{j}}_n}

\begin{document}

\preprint{AIP/123-QED}

\title{A complete quasiclassical map for the dynamics of interacting fermions}

\author{Amikam Levy}
\email{amikamlevy@gmail.com}
\affiliation{Department of Chemistry,  University of California, Berkeley, Berkeley, California 94720, United State}
\affiliation{The Raymond and Beverly Sackler Center for Computational Molecular and Materials Science, Tel Aviv University, Tel Aviv, Israel 69978
}
 
\author{
Wenjie Dou}%
\affiliation{Department of Chemistry,  University of California, Berkeley, Berkeley, California 94720, United State}

\author{Eran Rabani}
\email{eran.rabani@berkeley.edu}
\affiliation{Department of Chemistry,  University of California, Berkeley, Berkeley, California 94720, United State}
\affiliation{The Raymond and Beverly Sackler Center for Computational Molecular and Materials Science, Tel Aviv University, Tel Aviv, Israel 69978
}
\affiliation{Materials Sciences Division, Lawrence Berkeley National Laboratory, Berkeley, California 94720, United States}

\author{David T. Limmer}
\email{dlimmer@berkeley.edu}
\affiliation{Department of Chemistry,  University of California, Berkeley, Berkeley, California 94720, United State}
\affiliation{Materials Sciences Division, Lawrence Berkeley National Laboratory, Berkeley, California 94720, United States}
\affiliation{Kavli Energy NanoScience Institute, Berkeley, California 94720, United States}


\begin{abstract}
We present a strategy for mapping the dynamics of a fermionic quantum
system to a set of classical dynamical variables. The approach is
based on imposing the correspondence relation between the commutator
and the Poisson bracket, preserving Heisenberg's equation of motion
for one-body operators. In order to accommodate the effect of two-body
terms, we further impose quantization on the spin-dependent
occupation numbers in the classical equations of motion, with a
parameter that is determined self-consistently.  Expectation values
for observables are taken with respect to an initial
quasiclassical distribution that respects the original quantization of
the occupation numbers.  The proposed classical map becomes complete
under the evolution of quadratic Hamiltonians and is extended for all
even order observables.  We show that the map provides an accurate
description of the dynamics for an interacting quantum impurity model
in the coulomb blockade regime, at both low and high temperatures.  The
numerical results are aided by a novel importance sampling scheme that
employs a reference system to reduce significantly the sampling effort
required to converge the classical calculations.
\end{abstract}

\maketitle

\section{\label{sec:int}Introduction}

Molecular simulation is an indispensable tool for understanding many-body quantum systems driven away from equilibrium. 
Describing the dynamics of molecular- and mesoscopic-electronics on time-and length-scales relevant to experiments, however, is challenging.
In recent years,
significant progress has been made by introducing numerically
converged techniques, such as methods that rely on real-time
diagrammatic sampling
techniques\cite{rabani2008,weiss_iterative_2008,schiro_real-time_2009,werner_diagrammatic_2009,gull10_bold_monte_carlo,Segal10,cohen_memory_2011,hartle2013decoherence,Cohen-Gull-Reichman2015-introducing-inchworm},
wave function-based approaches such as numerical renormalization group
techniques
\cite{schmitteckert_nonequilibrium_2004,anders_real-time_2005,bulla2008numerical}
and multi-layer multiconfiguration methods,\cite{wang2018multilayer,balzer2015multiconfiguration} or reduced and
hierarchical density matrix approaches.\cite{hartle2013decoherence,schinabeck2016hierarchical} While significant
progress has been made using these methods to understand the transport
in various correlated
scenarios,\cite{wilner_bistability_2013,wilner_phonon_2014,wilner2014}
their application to more realistic systems is still 
limited. 

An alternative approach to these numerically converged techniques is
based on approximate methods that are more flexible in describing
realistic complex scenarios, but often introduce simplifications leading
to uncontrolled errors.  Among the more poplar methods are
master equations (QME) and their
generalizations,\cite{Datta1990,Mukamel2006,Leijnse2008,Esposito2009,Esposito2010,dou2015,arrigoni15}
and approaches based on the nonequilibrium Green's function methods
with specific closures for the
self-energy.\cite{hettler1998,Datta2000,Xue2002,Galperin07,Haug2008,stefanucci_nonequilibrium_2013}
More recently, quasiclassical mapping
techniques~\cite{Mayer1979,miller1986,van2004semiclassical,miller2006,swenson2011,swenson2012,li2012,li2013,li2014,davidson2017,montoya2018exact}
have been developed that 
cast the many-body quantum problem onto a set of classical dynamical variables and describe the transport in extended systems coupled
to complex non-linear environment, with varying coupling strengths.
Such classical mapping procedures further admit the use of advanced sampling
techniques of rare fluctuations\cite{ray2018importance} developed for classical molecular
dynamics simulations.

Previous attempts to map the dynamics of fermionic systems onto a set
of classical dynamical variables failed to reliably reproduce
correlation effects, such as the Coulomb blockade
staircase.\cite{swenson2011,li2012,li2013,li2014} This is mainly due
to the lack of quantization of the number operators in the classical
map, leading to a continuous increase of the current with the increase of
bias or gate voltage, in a quantum point-contact setup.  Moreover, the
description of the dynamics of observables that depend non-linearly on
a pair of creation and annihilation operators, for example, in
shot-noise measurements, has not received any attention. As shown
below, a naive and straightforward application of the classical maps
to such observables leads to significant errors, even for
noninteracting model Hamiltonian, where the map generates the exact
dynamics.  

In this study, we develop a new strategy to map the dynamics of an open
quantum system driven away from equilibrium onto a set of classical
dynamical variables.  The method maps a pair of creation or annihilation
fermionic operators to phase-space variables in Cartesian coordinates
that satisfies a correspondence relation between the commutator and
the Poisson brackets. 
In order to accommodate the
effect of two-body terms (electron-electron interactions), we further
impose quantization rules on the spin-dependent occupation numbers in
the classical equations of motion, with an onset parameter that is
determined self-consistently.
Combining this map
with the initial value
representation~\cite{herman1984,kay1994,stock1997} that incorporates
the discrete nature of quantum mechanics results in a robust
description of the dynamics on diverse time-scales, as illustrated for
the Anderson impurity model~\cite{Anderson1961} 
for a wide range of temperatures and on-site
electron-electron repulsion term.  We further show that for quadratic
Hamiltonians, higher order fermionic operators can be mapped accurately
as a consequence of completeness, providing a framework to study the
fluctuations and high order correlations within this mapping approach.
Finally, we develop a reference sampling approach to reduce
significantly the number of
trajectories required to converge expectation values.

\section{\label{sec:anderson}Anderson impurity model}
For concreteness, throughout this manuscript, we consider the evolution
of observables for the Anderson impurity model.  This model is defined
by the Hamiltonian $H=H_{S}+H_{B}+V$, where \beqar
\label{eq:HS}
\Hop_{S}= & \underset{\sigma=\uparrow,\downarrow}{\sum}
\varepsilon_{\sigma}
\dop_{\sigma}^{\dagger}\dop_{\sigma}+U\dop_{\uparrow}^{\dagger}\dop_{\uparrow}\dop_{\downarrow}^{\dagger}\dop_{\downarrow}
\eeqar describes the impurity (or dot), referred to simply as the
`system Hamiltonian', \beqar
\label{eq:HB}
\Hop_{B}= & \underset{\underset{k\in \rm L,\rm R}{ \sigma=\uparrow,\downarrow}}{\sum}\varepsilon_{k}\cop_{k\sigma}^{\dagger}\cop_{k\sigma}
\eeqar
describes the noninteracting fermionic baths (or leads), and 
\beqar
\label{eq:V}
\hat{V}= & \underset{\underset{k\in
    \rm L,\rm R}{\sigma=\uparrow,\downarrow}}{\sum}
t_{k}\dop_{\sigma}^{\dagger}\cop_{k\sigma}+{\rm h.c.}, \eeqar
describes the hybridization between the system and the leads.  Here,
$d_{\sigma}^{\dagger}\ \left(d_{\sigma}\right)$ are the creation
(annihilation) operators of an electron on the dot with spin
$\sigma=\uparrow,\downarrow$ with a one-body energy
$\varepsilon_{\sigma}$.  $U$ is the on-site Hubbard interaction,
$\cop_{k\sigma}^{\dagger}\ \left(\cop_{k\sigma}\right)$ are the creation
(annihilation) operators of an electron in mode $k$ of the leads with
energy $\varepsilon_{k}$, and $t_{k}$ is the hybridization between the
dot and mode $k$ in the lead.  The coupling to the quasi-continuous
leads is modeled in the wide band limit.  The spectral function of the
left ($\ell=\rm L$) or right ($\ell=\rm R$) lead is:
\beq
J_\ell(\varepsilon_k)=\frac{\Gamma_\ell
}{\left(1+e^{A(\varepsilon_k-B/2}
  \right)\left(1+e^{-A(\varepsilon_k+B/2} \right)}.
\eeq
where $\Gamma_\ell$ determines the coupling strength to the
$\ell$-lead, $B$ is the width of the spectral function, and $A$
determines the sharpness of the cutoff.  The coupling $t_k$ between
the dot and the $k$-th mode is expressed in terms of the spectral
function as
$t_{k\in\ell}=\sqrt{J_{\ell}\left(\ensuremath{\varepsilon}_{k}\right)\Delta\epsilon/2\pi}
$, where $\Delta\varepsilon=2\varepsilon_\mathrm{max}/ (N_{\ell}/2
-1)$ is the discretization of the leads energy spectrum, $N_{\ell}$ is
the numbers of modes in the $\ell$-lead, and
$2\varepsilon_\mathrm{max}$ is energy range in the leads.  To model
accurately the wide band limit, one should consider sufficiently large
values for $B$ such that the energy scale of the system is encompassed
inside the spectrum of the leads and that the modes in the leads are
dense enough, i.e.  $\Delta\varepsilon$ is sufficiently small.  In the
simulations below each lead consists of $N_{\ell}=600$ modes, where
half are with spin up and the other half with spin down.  Throughout,
we take $\hbar$, $k_B$ and the charge of the electron $e$, to be 1.

\begin{figure}[t]
\center{\includegraphics[width=8cm]{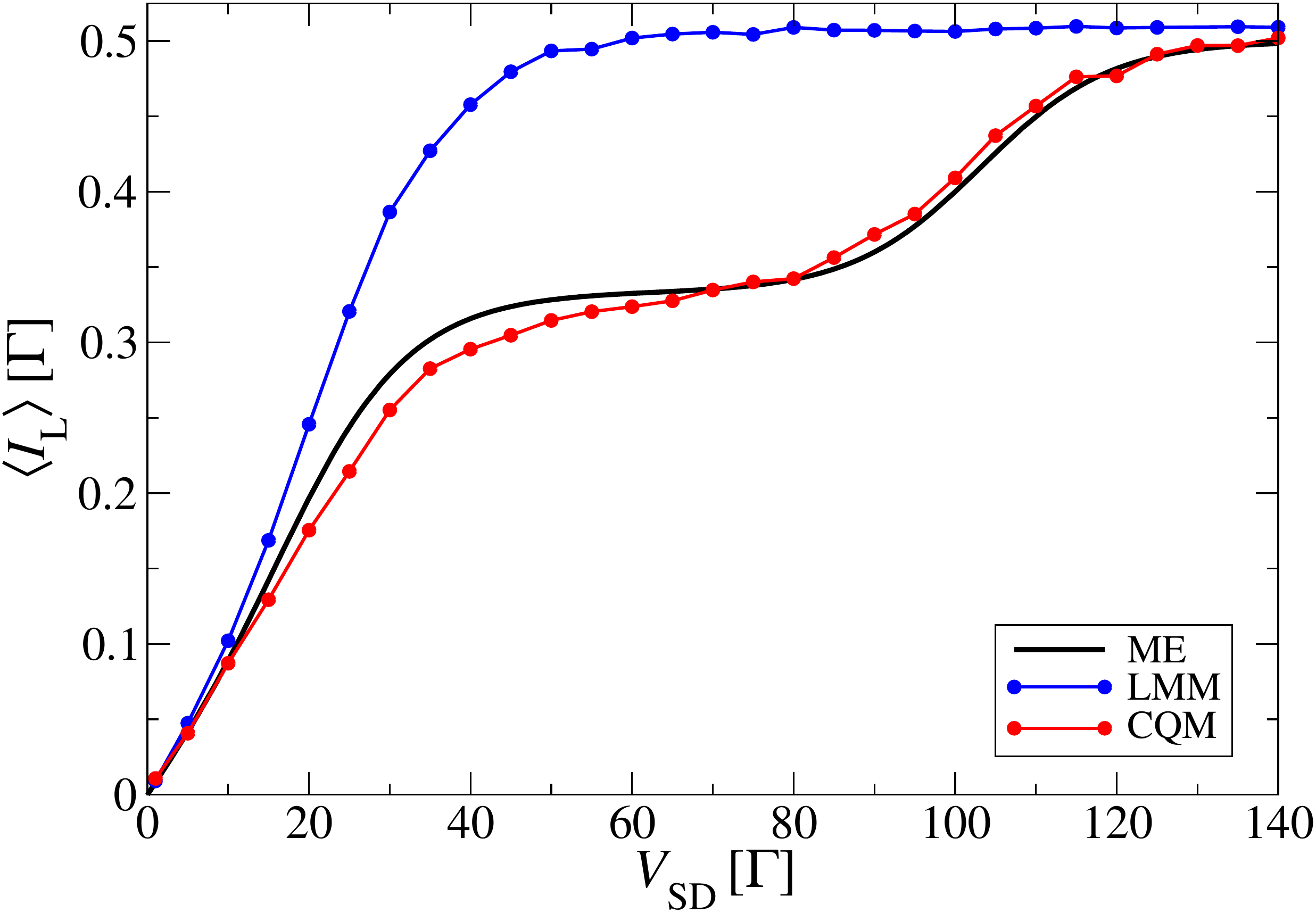}}
\caption{The steady-state current from the left lead ($\langle I_L
  \rangle$) as function of the bias voltage ($V_{\rm SD}$). The black
  symbols are the results from the QME approach. The blue and red
  curves correspond to the LMM and CQM, respectively.  Parameters
  used: $\Gamma=2\Gamma_{\rm L}=2\Gamma_{\rm R}=1$,
  $\epsilon_{\uparrow}=\epsilon{\downarrow}=10\Gamma$, $T=4\Gamma$,
  $U=40\Gamma$, $\Delta_{\sigma}=0.32$, $N_{\rm tr}=3\times10^4$, and
  $\mu_{\rm L}=-\mu_{\rm R}=V/2$.  }
\label{fig:I_V}
\end{figure}

To assess the accuracy and robustness of the quasiclassical mapping
procedure, we focus on the Coulomb blockade effect that is manifested
by a staircase structure of the current versus voltage, as shown in
Fig.~\ref{fig:I_V}.  When the bias voltage is not sufficiently large
to overcome the on-site repulsion energy, only one conductance channel
is open. When the bias becomes sufficiently large compared to $U$, an
additional conducting channel opens up, and the current increases to
its maximal value of a two-channel quantum point-contact.  In
Fig.~\ref{fig:I_V}, we show the results of two quasiclassical mapping
procedures. The mapping approach that is isomorphic to quaternions
(Li-Miller map (LMM))~\cite{li2012} provides an accurate description
of the I-V characteristics at low and high bias voltages ($V_{\rm
  SD}$), but fails to reproduce the Coulomb blockade staircase. On the
other hand, the current complete quasiclassical map (CQM) provides a
qualitative description across all values of $V_{\rm SD}$. In
particular, it captures the staircase structure characteristic of the
Coulomb Blockade effect.  We will return to discuss the CQM results
after we introduce the strategy of mapping quantum to classical
degrees of freedom.

\section{\label{sec:map}Complete quasiclassical map (CQM)}
For an operator $\hat{A}$, in the Hilbert space of the Anderson
impurity model, the Heisenberg equation of motion reads
\beq
\label{eq:eq_motion}
\frac{d\hat{A}}{dt}= i\left[\Hop_0 , \hat{A} \right] +
iU\left[\nop_{\uparrow}, \Aop \right]\nop_{\downarrow} +
iU\nop_{\uparrow} \left[\nop_{\downarrow}, \Aop \right].  \eeq where
$\Hop_0$, \beq \Hop_0= \underset{\sigma=\uparrow,\downarrow}{\sum}
\varepsilon_{\sigma} \dop_{\sigma}^{\dagger}\dop_{\sigma}
+\underset{\underset{k\in \rm L,\rm R}{
    \sigma=\uparrow,\downarrow}}{\sum}\varepsilon_{k}\cop_{k\sigma}^{\dagger}\cop_{k\sigma}
+ \underset{\underset{k\in \rm L, \rm R}{\sigma=\uparrow,\downarrow}}{\sum}
t_{k}\dop_{\sigma}^{\dagger}\cop_{k\sigma}+{\rm h.c.}
\eeq
is the one-body, noninteracting part of the Hamiltonian. We wish to
find a map for $\Aop$ to a function of classical phase-space
variables, $A[\vec{R}]$, that would preserve the dynamics $\ave{\Aop
  (t)}=\ave{A (t)}_c$, where $\ave{\Aop(t)}= \mathrm{Tr}
\left(\hat{\rho} \Aop(t) \right)$, and
\beq
\label{eq:calssical_average}
\ave{A(t)}_c = \int d \vec{R}\, \rho_0(\vec{R}) A[\vec{R(t)}]
\eeq
is the classical expectation value with respect to the initial
probability distribution $\rho_0$ of the total system.  We do this in
two parts. First we construct a complete map for quadratic
Hamiltonians, which is extendable to any even order operator, and valid for
noninteracting fermionic systems. Then, we propose a strategy for
mapping Hamiltonians of higher order containing onsite Hubbard
interactions.
In all the simulations, the equations of motion are solved numerically using an adaptive Rung-Kutta(4,5) method.
The number of trajectories, $N_{\rm tr}$, used to converge the results is specified below for each case study. For the steady-state results, additional time averaging is considered.
\subsection{\label{subsec:non-int}Noninteracting fermions}
We first consider the case of non-interacting fermions, $U=0$,
described by a Hamiltonian that depends quadratically on the creation
and annihilation operators, $\Hop_0$.  To reproduce the dynamics of
the expectation value of quadratic operators under the evolution
describe in Eq.~(\ref{eq:eq_motion}) we require:
\begin{enumerate}[label=(\alph*)]
\item For any quadratic operator $\hat{A}$, and its classical
  counterpart $A$, the commutator and the Poisson bracket satisfies
  the correspondence relation $i\left[\Hop_0 , \hat{A} \right] =
  \left\lbrace A , H_0 \right\rbrace$.
\item For any quadratic expectation value, the initial probability
  distribution $\rho_0$ must satisfy $\ave{\hat{A}(0)}=\ave{A(0)}_c$,
  and respect the quantum discrete nature of the occupations.
\end{enumerate}
It is straightforward to show that condition (a) is satisfied by
mapping a pair of creation and annihilation operators to a phase space
of conjugated variables,
$\vec{R}=({x},{p}_{x},{y},{p}_{y})$, as \beqar
\label{eq:quad_crea/anni_map}
\aop_n^{\dagger}\aop_n &\mapsto& x_n p_{y,n}- y_n p_{x,n} ,\\ \nonumber
\aop_n^{\dagger}\aop_m &\underset{m\neq n}{\mapsto}& \frac{1}{2}\left [ i\left(x_n p_{x,m} - p_{x,n} x_m + y_n p_{y,m} - p_{y,n} y_m \right) \right . \\ \nonumber
&&+\left . \left(x_n p_{y,m} - p_{x,n} y_m + x_m p_{y,n} - p_{x,m} y_n \right) \right ] , \\ \nonumber
\aop_n^{\dagger}\aop_m^{\dagger} &\mapsto& \frac{1}{2}\left [ i\left(x_n p_{x,m} - p_{x,n} x_m - y_n p_{y,m} + p_{y,n} y_m \right) \right .\\ \nonumber
&&- \left . \left(x_n p_{y,m} - p_{x,n} y_m - x_m p_{y,n} + p_{x,m} y_n \right) \right ] ,\\ \nonumber
\aop_n\aop_m &\mapsto& \frac{1}{2}\left [i\left(x_n p_{x,m} - p_{x,n} x_m - y_n p_{y,m} + p_{y,n} y_m \right) \right . \\ \nonumber
&&+ \left .  \left(x_n p_{y,m} - p_{x,n} y_m - x_m p_{y,n} + p_{x,m} y_n \right) \right ] \, , \\ \nonumber
\eeqar  
This identifies positions, $(x,y)$ and their conjugate momenta,
$({p}_x,{p}_y)$, and the Poisson bracket can be used to check that
this map returns the quantum commutator of any pair of quadratic
creation/annihilation operators. Because any quadratic Hamiltonian
with a set of quadratic operators constitutes a closed Lie algebra of
quadratic operators, condition (a) insures a loyal representation of
the dynamics in terms of Hamilton's equations.  We note in passing
that we subtracted $1/2$ from the classical map of
$\nop_i\equiv\aop_i^{\dagger}\aop_i$ to include a Langer-like
correction\cite{miller1978classical}.

For leads that are in thermal equilibrium and uncorrelated initial
state, condition (b) can be satisfied by setting the initial
occupation of each mode in the left and right leads to a value 0 or 1,
such that the expectation value, averaged over the set of initial
conditions, satisfies the Fermi-Dirac distribution.\cite{swenson2011}
Operationally, we choose a random number $\xi_{k\sigma}\in [0,1]$ and
then select the occupation of mode $k\sigma$ of the $\ell$-lead
according to
\beq
\label{eq:initial_0ccupation}
    n_{k\sigma}=\left\{
                \begin{array}{ll}
                  0 \quad \xi_{k\sigma}> \left(1+e^{\beta_{\ell}(\epsilon_k-\mu_{\ell})}\right)^{-1}\\
                  1 \quad \xi_{k\sigma}\leq \left(1+e^{\beta_{\ell}(\epsilon_k-\mu_{\ell})}\right)^{-1} 
                 \end{array}
\right. ,              
\eeq
where $\beta_{\ell}=1/k_{\rm B} T_{\ell}$ and $\mu_l$ are the inverse
temperature times Boltzmann's constant and chemical potential of the
$\ell$-lead, respectively.  The Cartesian coordinates are then sampled
according to~\cite{li2013}
\beqar
\label{eq:sample}
x_{k\sigma}&=& \cos(\theta_{k\sigma}), \quad
p_{x,n\sigma}=-n_{k\sigma}\sin(\theta_{k\sigma}) \\ \nonumber
y_{k\sigma}&=& \sin(\theta_{k\sigma}), \quad p_{y,n\sigma}=
n_{k\sigma}\cos(\theta_{k\sigma}), \eeqar where $\theta_{k\sigma}$ is
chosen randomly in the interval $[0,2\pi]$ and
$n_{k\sigma}=x_{k\sigma}p_{y,k\sigma}-y_{k\sigma}p_{x,k\sigma}$
satisfies Eq.~(\ref{eq:initial_0ccupation}), resulting in
$\ave{\nop_{k\sigma}}=\ave{n_{k\sigma}}_c=(1+\exp[\beta_{\ell}(\epsilon_k-\mu_{\ell})])^{-1}
$ at the initial time.  By construction, the expectation value at the
initial time is $\ave{a_n^{\dagger} a_m}_c=\ave{a_n^{\dagger}
  a_m^{\dagger}}_c=\ave{a_n a_m}_c =0$, as expected for averages taken
with respect to uncorrelated thermal distribution.  The sampling
choice in Eq.~(\ref{eq:sample}) is not unique, but it does provide an
efficient averaging of the expectation values with respect to the
number of trajectories.\cite{li2013} In a similar manner, one can set
the initial occupation of the dot.

Comparing the proposed CQM given by Eq.~(\ref{eq:quad_crea/anni_map})
to the LMM, we find that the mappings of the diagonal term
$\aop_n^{\dagger}\aop_n$ and of the linear combination
$\aop_n^{\dagger}\aop_m^{\dagger}+\aop_m^{\dagger}\aop_n^{\dagger}$
are identical in both maps, but the remaining terms in
Eq.~(\ref{eq:quad_crea/anni_map}) cannot be expressed using the LMM.
This leads to the Hamiltonian being expressed identically in both maps,
\beqar
\label{eq:H0}
H_0 &=& \underset{\sigma=\uparrow,\downarrow}{\sum} \varepsilon_{\sigma}\left(x_{\sigma}p_{y,\sigma}-y_{\sigma}p_{x,\sigma} \right)\\ \nonumber
&+&\underset{\underset{k\in L,R}{ \sigma=\uparrow,\downarrow}}{\sum}\varepsilon_{k} \left(x_{k\sigma}p_{y,k\sigma}-y_{k\sigma}p_{x,k\sigma} \right)\\ \nonumber
&+&\underset{\underset{k\in L,R}{\sigma=\uparrow,\downarrow}}{\sum} t_{k} \left( x_{\sigma}p_{y,k\sigma} - y_{\sigma}p_{x,k\sigma} + x_{k\sigma}p_{y,\sigma} - y_{k\sigma}p_{x,\sigma} \right) \,.
\eeqar
For this non-interacting Hamiltonian, mapping $H_0$ and then deriving
Hamilton's equations of motion for the phase space variables is
identical to deriving Heisnberg's equation of motion for the bi-linear
operators and then mapping the results using
Eq.~(\ref{eq:quad_crea/anni_map}). 

We can also map other quadratic observables, such as the current from
the left lead:
\beqar
\Iop_{\rm L} &&=-\frac{d}{dt} \underset{\underset{k\in \rm L}{
    \sigma=\uparrow,\downarrow}}{\sum}\cop_{k\sigma}^{\dagger}\cop_{k\sigma}
\mapsto \\ \nonumber && \underset{\underset{k\in \rm L}{
    \sigma=\uparrow,\downarrow}}{\sum} t_k\left(y_{\sigma}
p_{y,k\sigma}-p_{y,\sigma} y_{k\sigma} + x_{\sigma}
p_{x,k\sigma}-p_{x,\sigma} x_{k\sigma} \right) \, .
\eeqar
As a diagonal term, the above form is also identical to the expression
obtained by the LMM.\cite{li2013} In the upper panel of
Fig.~\ref{fig:II} we compare the results for the left current
generated by the CQM (which in this case are equivalent to the LMM)
with exact quantum mechanical results for a noninteracting model
Hamiltonian.  As expected, the agreement between the CQM (or the LMM)
and exact quantum mechanical results is excellent.  
In the next section we show that for the CQM  these result can be extended to higher-order operators.

\subsection{\label{subsec:HigerOrder}Higher order operators}
Mapping higher order operators, operators that involve more
than one pair of creation/annihilation operators, is more difficult
due to the nonlocal character of fermions arising from their exclusion
statistics.  Ignoring the fermionic nature does not seem to make any
significant difference for a single pair of creation/annihilation
operators,\cite{swenson2011,li2013} but for higher order operators,
the anti-commutation of the creation/annihilation fermionic operators
plays a significant role and describing quantum fluctuations such as
shot noise requires a careful consideration of this effect.

For example, consider a map for the operator $\hat{A}=a_n^\dagger
  a_m a_m^\dagger a_k$. Using the anti-commutation nature of
$\{a_n^\dagger,a_n\}=1$, we can express the expectation value of
$\hat{A}$ using four different terms that are identical quantum
mechanically, but differ when mapped onto classical phase space
variables. Specifically, expanding $\ave{\hat{A}}$
\beqar
\label{eq:representation}
\ave{\aop_n^{\dagger} \aop_m \aop_m^{\dagger} \aop_k }=&& C_1\ave{ \left( \aop_n^{\dagger}\aop_m  \right) \left(\aop_m^{\dagger} \aop_k \right)} \\ \nonumber
&+&C_2\ave{\left(\aop_n^{\dagger}\aop_k\right) -\left(\aop_n^{\dagger}\aop_m^{\dagger}\right)\left(\aop_m\aop_k\right) } \\ \nonumber
&+& C_3\ave{\left(\aop_n^{\dagger}\aop_k\right)\left(\aop_m\aop_m^{\dagger}\right) } \\ \nonumber
&+&C_4\ave{\delta_{nk}\left(\aop_m\aop_m^{\dagger}\right)- \left(\aop_k\aop_m\right)\left(\aop_m^{\dagger}\aop_n^{\dagger}\right)}. 
\eeqar  
we find there are four unique combinations of operators, which generically have coefficients, $C_i$. To determine the best choice of $C_i$, we impose conditions (a) and
(b) of Sec.~\ref{subsec:non-int} on the time evolution of $\hat{A}$
and require that the time evolution of $\hat{A}$ be exact for a
quadratic Hamiltonian, i.e., that $i[\Hop,\Aop] = \lbrace A,H \rbrace$ and that
$\ave{\Aop(0)}=\ave{A(0)}_c$.  For an uncorrelated initial thermal
state, the values that satisfy these conditions are  $C_1=1, C_2=-1, C_3=1$, and $C_4=0$. Note that
Eq.~(\ref{eq:representation}) contains pairs of creation or
annihilation operators $(a_n^\dagger a_m^\dagger$ or $a_n a_m)$, which
cannot be described within the LMM.

\begin{figure}[t]
\center{\includegraphics[width=8cm]{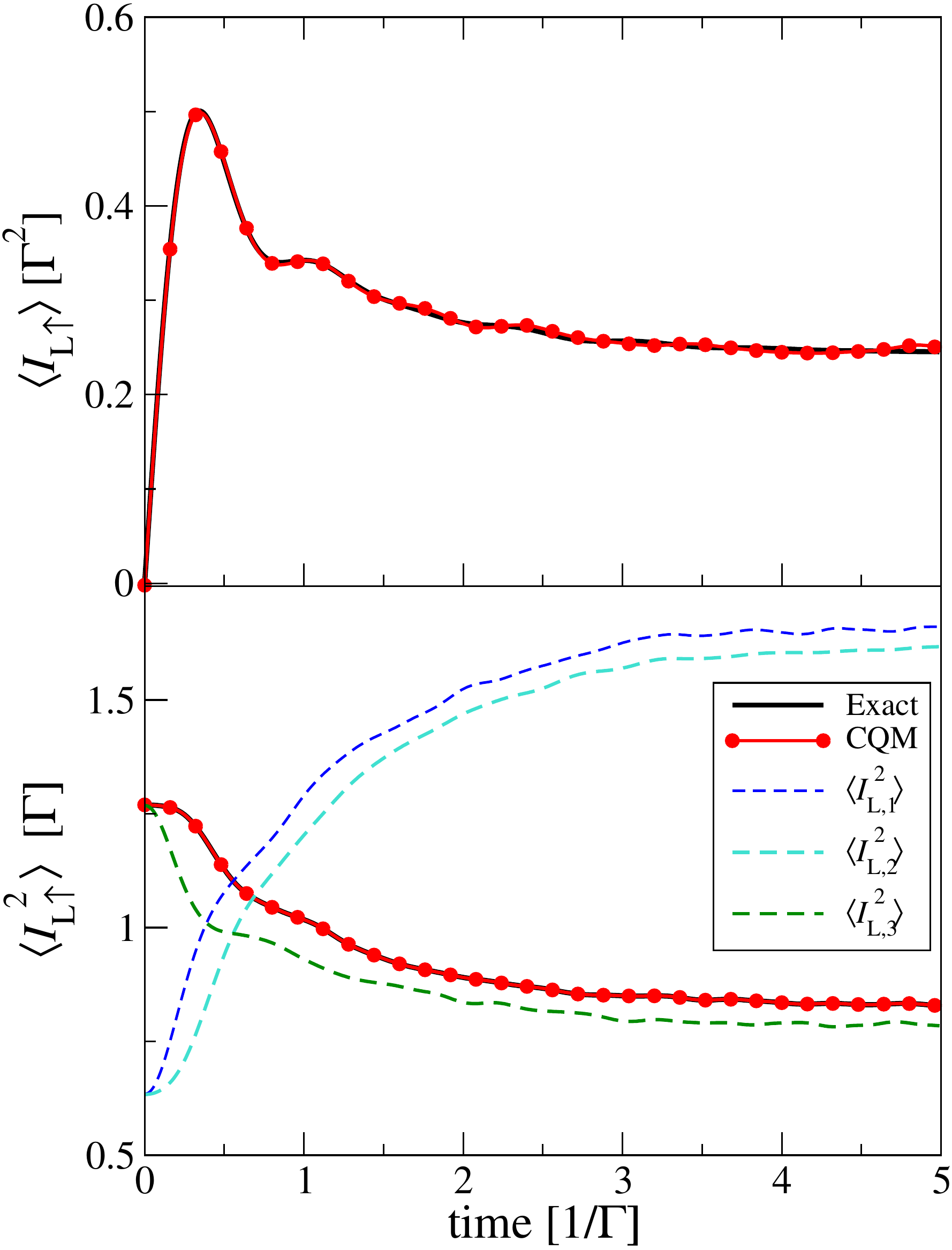}}
\caption{Upper panel: The average left current for spin up ($\langle
  I_{\rm L\uparrow}(t) \rangle$) as a function of time for a
  non-interacting model Hamiltonian. The solid black line represents
  the exact quantum mechanical result~\cite{swenson2011} and a the red
  symbols are the result of CQM. Lower panel: The average left current
  squared for spin up ($\langle I^2_{\rm L\uparrow}(t) \rangle$) as a
  function of time. In addition to the exact quantum mechanical (solid
  black curve) and CQM (red symbols) results, we also show the
  individual terms $\langle I^2_{\rm L,1}(t) \rangle$ (blue), $\langle
  I^2_{\rm L,2}(t) \rangle$ (cyan), and $\langle I^2_{\rm L,3}(t)
  \rangle$ (green).  Parameters used: $\Gamma=2\Gamma_{\rm
    L}=2\Gamma_{\rm R}=1$,
  $\epsilon_{\uparrow}=\epsilon{\downarrow}=-\Gamma$,
  $T=\frac{\Gamma}{5}$, $U=0$, $N_{\rm tr}=10^5$, and $\mu_{\rm
    L}=-\mu_{\rm R}=6\Gamma$.}
\label{fig:II}
\end{figure} 

Applying this procedure to the second moment of the left current for
the noninteracting Hamiltonian yields a simple expression for
$\ave{\Iop_{\rm L}^2}=\ave{\Iop_{\rm L,1}^2}-\ave{\Iop_{\rm
    L,2}^2}+\ave{\Iop_{\rm L,3}^2}$, where
\beqar
\label{eq:II}
\\\nonumber
\ave{\Iop_{\rm L,1}^2}&=&
    \underset{\underset{j,k\in L}{ \sigma=\uparrow,\downarrow}}{\sum}
    t_{j}t_{k}\left\langle
    \left(\cop_{j\sigma}^{\dagger}\dop_\sigma\right)\left(\dop_{\sigma}^{\dagger}\cop_{k\sigma}\right)
    +\left(\dop_{\sigma}^{\dagger}\cop_{j\sigma}\right)\left(\cop_{k\sigma}^{\dagger}\dop_{\sigma}\right)\right\rangle\\ \nonumber
\ave{\Iop_{\rm L,2}^2}&=& \underset{\underset{j,k\in L}{ \sigma=\uparrow,\downarrow}}{\sum}      t_{j}t_{k}\left\langle \left(\cop_{j\sigma}^{\dagger}\cop_{k\sigma}\right)-\left(\cop_{j\sigma}^{\dagger}\dop_{\sigma}^{\dagger}\right)\left(\dop_{\sigma}\cop_{k\sigma}\right) \right .  \\ \nonumber
&&+
\left. \delta_{jk}\left(\dop_{\sigma}^{\dagger}\dop_{\sigma}\right)-\left(\dop_{\sigma}^{\dagger}\cop_{k\sigma}^{\dagger}\right)\left(\cop_{j\sigma}\dop_{\sigma}\right)\right\rangle
\\ \nonumber
\ave{\Iop_{\rm L,3}^2}&=& \underset{\underset{j,k\in L}{ \sigma=\uparrow,\downarrow}}{\sum} t_{j}t_{k}\left\langle \left(\cop_{j\sigma}^{\dagger}\cop_{k\sigma}\right)\left(\dop_{\sigma}\dop_{\sigma}^{\dagger}\right) +\left(\dop_{\sigma}^{\dagger}\dop_{\sigma}\right)\left(\cop_{j\sigma}\cop_{k\sigma}^{\dagger}\right)\right\rangle.  \\ \nonumber
\eeqar
As can be shown explicitly, this mapping of the second moment of the left current operator
satisfies both conditions (a) and (b).

\begin{figure*}[t]
\center{\includegraphics[width=13cm]{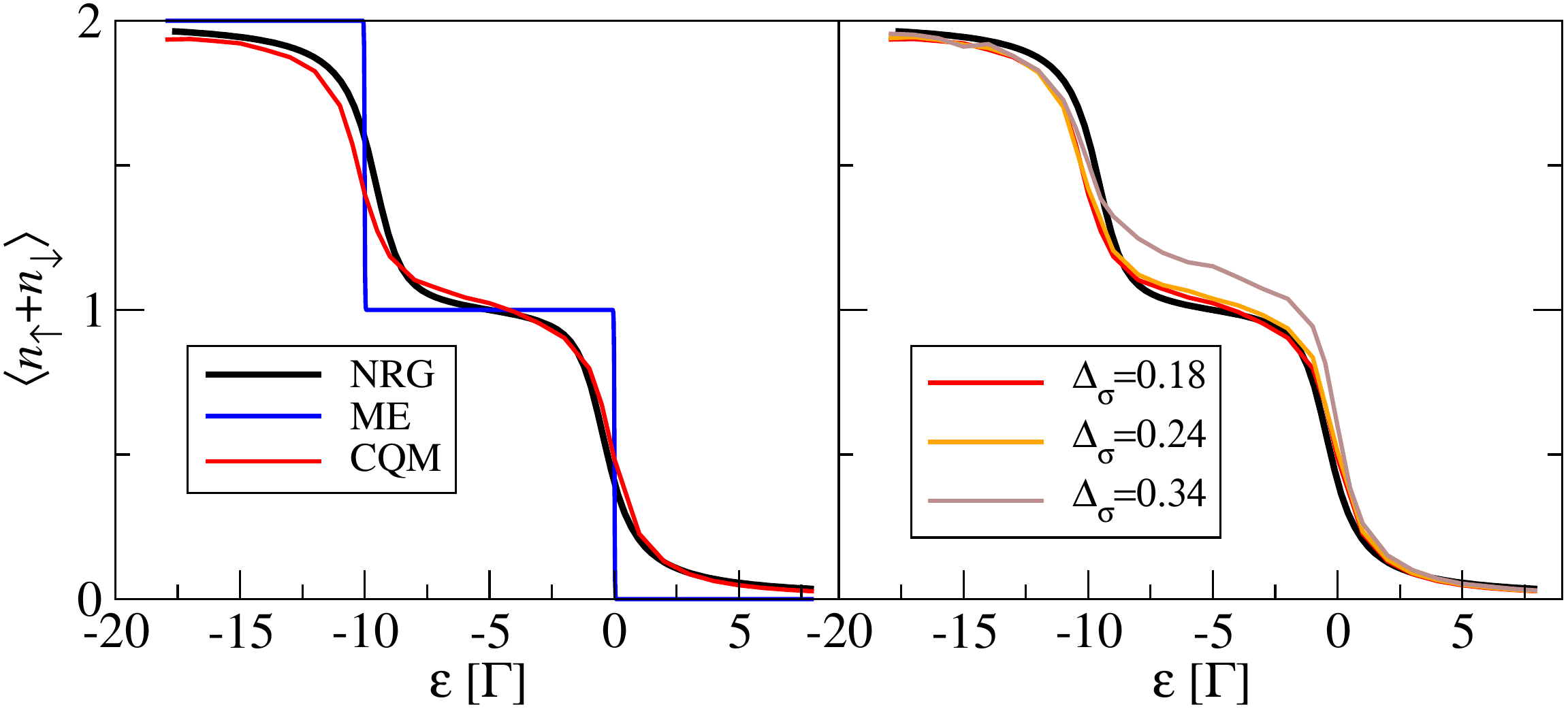}}
\caption{Left panel: The steady-state quantum dot population as a
  function of the gate voltage
  $\varepsilon=\varepsilon_{\uparrow}=\varepsilon_{\downarrow}$ under
  equilibrium conditions ($\mu_L=\mu_R=0$).  Right panel: The
  dependence of the steady-state dot population on the choice of the
  value of $\Delta_\sigma$ (cf., Eq.~(\ref{eq:map_interacting})).
  Parameters used: $\Gamma=2\Gamma_L=2\Gamma_R=1$, $T=\Gamma/100$,
  $U=10\Gamma$, $N_{\rm tr}=3\cdot10^4$, and for the left panel: $\Delta_{\sigma}=0.18$.}
\label{fig:n_epsilon}
\end{figure*}

In the bottom panel of Fig.~\ref{fig:II} we show the time evolution of
$\ave{\Iop_{\rm L}^2}$ for a quadratic Hamiltonian where $U=0$. The
agreement between the exact quantum mechanical result and the CQM is
excellent. We also plot the individual terms $\ave{\Iop_{\rm L,1}^2}$,
$\ave{\Iop_{\rm L,2}^2}$, and $\ave{\Iop_{\rm L,3}^2}$ (dashed lines).  Only
the proper combination of all three terms yields an accurate
description of $\ave{\Iop_{\rm L}^2}$.  We note that the LMM
can only be used to map the first term, but not the other two that
contribute to $\ave{\Iop_{\rm L}^2}$.

\subsection{\label{subsec:int}Interacting fermions}
The on-site interaction, $U \nop_{\uparrow} \nop_{\downarrow}$, that
manifest the Coulomb blockade effect, is a four index term
that is outside the space defined by the CQM. In order to account for
this two-body interaction term, we map the two terms proportional to
$U$ in Eq.~(\ref{eq:eq_motion}) according to
\beqar
\label{eq:map_interacting}
iU\left[\nop_{\uparrow}, \Aop \right]\nop_{\downarrow} \mapsto U\left\lbrace A_c , n_{\uparrow } \right\rbrace \theta\left(n_{\downarrow }-\Delta_{\downarrow} \right) \\ \nonumber
iU \nop_{\uparrow} \left[\nop_{\downarrow}, \Aop \right] \mapsto U\left\lbrace A_c , n_{\downarrow } \right\rbrace \theta\left(n_{\uparrow }-\Delta_{\uparrow} \right),
\eeqar 
where $\theta$ is the Heaviside step function.  The idea behind this
choice is that the term $U\nop_{\uparrow} \nop_{\downarrow}$
contributes to the dynamics only when both electrons with spin up and
spin down occupy the site.  Classically, the occupation number
admits a continuous value, which implies that the Hubbard term can
become significant for fractional populations of the two
spin-channels.  Much like a mean-field approximation, these fraction
contributions of the Hubbard term will smear the Coulomb blockade
effect.  By introducing the step function, we impose that
contributions to the dynamics from the Hubbard term arise only in
trajectories for which
$n_{\uparrow(\downarrow)}>\Delta_{\uparrow(\downarrow)}$.  The
parameter $\Delta_{\uparrow(\downarrow)}$ is determined according to
the distribution of $n_{\uparrow(\downarrow)c}$ and will be discussed
in detail below. 
We note that the classical expression in Eq.~(\ref{eq:map_interacting}) is not derivable from a Hamiltonian, and therefore does not in principle conserve energy or the norm of phase space. Nevertheless, we find relaxation to an intermediate time, long lived steady-state for all of the observables studied on timescales shorter than the recurrence times. 

\par
Considering the Anderson impurity model, the equations of motion for
the Cartesian variables for the lead degrees of freedom are
\beqar
\dot{x}_{j\sigma} &=&-\epsilon_{j}y{}_{j\sigma}-t_{j}y_{\sigma},
\\ \nonumber
\dot{y}_{j\sigma} &=&\epsilon_{j}x{}_{j\sigma}+t_{j}x_{\sigma},\\ \nonumber
\dot{p}_{x,j\sigma} &=&-\epsilon_{j}p_{y,j\sigma}-t_{j}p_{y,\sigma},\\ \nonumber
\dot{p}_{y,j\sigma} &=&\epsilon_{j}p_{x,j\sigma}+t_{j}p_{x,\sigma},\\ \nonumber
\eeqar
and those for the system's degrees of freedom are
\beqar
\dot{x}_{\sigma}&=&-\epsilon_{\sigma}y_{\sigma}-\sum_{k\in L,R}t_{k}y_{k\sigma} - Uy_{\sigma}\theta\left(n_{\bar{\sigma}} -\Delta_{\bar{\sigma}} \right), \\ \nonumber
\dot{y}_{\sigma}&=&\epsilon_{\sigma}x_{\sigma}+\sum_{k\in L,R}t_{k}x_{k\sigma} + Ux_{\sigma}\theta\left(n_{\bar{\sigma}} -\Delta_{\bar{\sigma}} \right), \\ \nonumber
\dot{p}_{x,\sigma}&=&-\epsilon_{\sigma}p_{y,\sigma}-\sum_{k\in L,R}t_{k}p_{y,k\sigma} - Up_{y,\sigma}\theta\left(n_{\bar{\sigma}} -\Delta_{\bar{\sigma}} \right), \\ \nonumber
\dot{p}_{y,\sigma}&=&\epsilon_{\sigma}p_{x,\sigma}+\sum_{k\in L,R}t_{k}p_{x,k\sigma} + Up_{x,\sigma}\theta\left(n_{\bar{\sigma}} -\Delta_{\bar{\sigma}} \right), \\ \nonumber
\eeqar
where $\bar{\sigma}=\downarrow,\uparrow$ is the opposite spin to $\sigma=\uparrow,\downarrow$.

In Fig.~\ref{fig:I_V} we plot the I-V curve obtained by the CQM and
compare it to the QME in Ref.~\onlinecite{levy19} and to the results obtained by the LMM.
We consider the limit of weak system-bath coupling and high
temperature where the QME provides a good approximation for the
dynamics of the system.  The LMM provides a good description of the
I-V characteristics at low and high bias voltages, but it fails to
capture the staircase structure reminiscent of the Coulomb blockade.
The CQM reproduce the QME results quantitatively, specifically, it
captures the staircase structure due to the Coulomb blockade effect. In
this high temperature regime, the agreement between the CQM and the
quantum mechanical results is observed for a wide range of onsite
Hubbard repulsion term and also for the quantum dot population.

Next, we consider a regime where the QME breaks down, namely, the low
temperature regime. For simplicity we focus on the equilibrium case
where $\mu_L=\mu_R=0$. In this regime, solutions for the population as
function of the gate voltage
($\varepsilon=\varepsilon_{\uparrow}=\varepsilon_{\downarrow}$) are
readily available using the numerical renormalization group (NRG)
technique.\cite{bulla2008,dou2017} In the left panel of Fig.~\ref{fig:n_epsilon} we
plot the quantum dot total population ($\langle n_\uparrow+n_\downarrow
\rangle$) as a function of the gate voltage.  The NRG results show a
staircase shape which is a manifestation of the Coulomb blockade
effect.  The CQM agrees quantitatively with the NRG results over a
wide range of gate voltages.  Specifically, it captures both the
position of the blockade as well as its width.  
The QME approach, however, captures only the position of the resonances; the broadening of the transitions are missing completely.  
This qualitative
difference between the CQM and QME approaches signifies the advantages
of the quasiclassical mapping techniques over the commonly used QME
approach for a broad range of temperatures.
\par
The mapping of the Hubbard term in Eq.~(\ref{eq:map_interacting})
introduces a parameter, $\Delta_{\uparrow(\downarrow)}$, which is
determined self-consistently. For the results show in
Fig.~\ref{fig:n_epsilon} we use a single value for
$\Delta_{\sigma}=0.18$, determined by considering the particle-hole
symmetry point, where
$\varepsilon_\uparrow=\varepsilon_\downarrow=-U/2$. At the symmetric
point, the steady state value of the average dot populations is
$\ave{\nop_{\sigma}}=1/2$.  Since quantum mechanically $\nop_{\sigma}$
can only assume two value, $0$ or $1$, whereas the distribution of
$n_{\sigma}$ is continuous, to obtain an average dot population
$\ave{\nop_{\sigma}}=1/2$, we set $\Delta_{\sigma}$ to the median of
the distribution of values of $n_{\sigma}$.  This ensures that the
Hubbard terms in Eq.~(\ref{eq:map_interacting}) are significant only when
$n_{\sigma}$ is sufficiently large, modeling the discrete nature of
the spin-dependent dot occupations.
\par
In the right panel of Fig.~\ref{fig:n_epsilon} we show the results obtained
for the total dot occupation for different values of
$\Delta_{\sigma}$. Only $3$ iterations are required to converge the
results for $\Delta_{\sigma}$.  We start with an initial guess of
$\Delta_{\sigma}=0.34$.  In the next iteration, we set
$\Delta_{\sigma}$ to the new median of the distribution of values of
$n_{\sigma}$, in this case $\Delta_{\sigma}=0.24$. We then repeat this
procedure until convergence (only one more iteration is required). The
results clearly show that the expectation value is not very sensitive to small
variations in the value of $\Delta_{\sigma}$, but the converged
results provide the best agreement with the NRG results.
\par
\begin{figure}[t]
\center{\includegraphics[width=8cm]{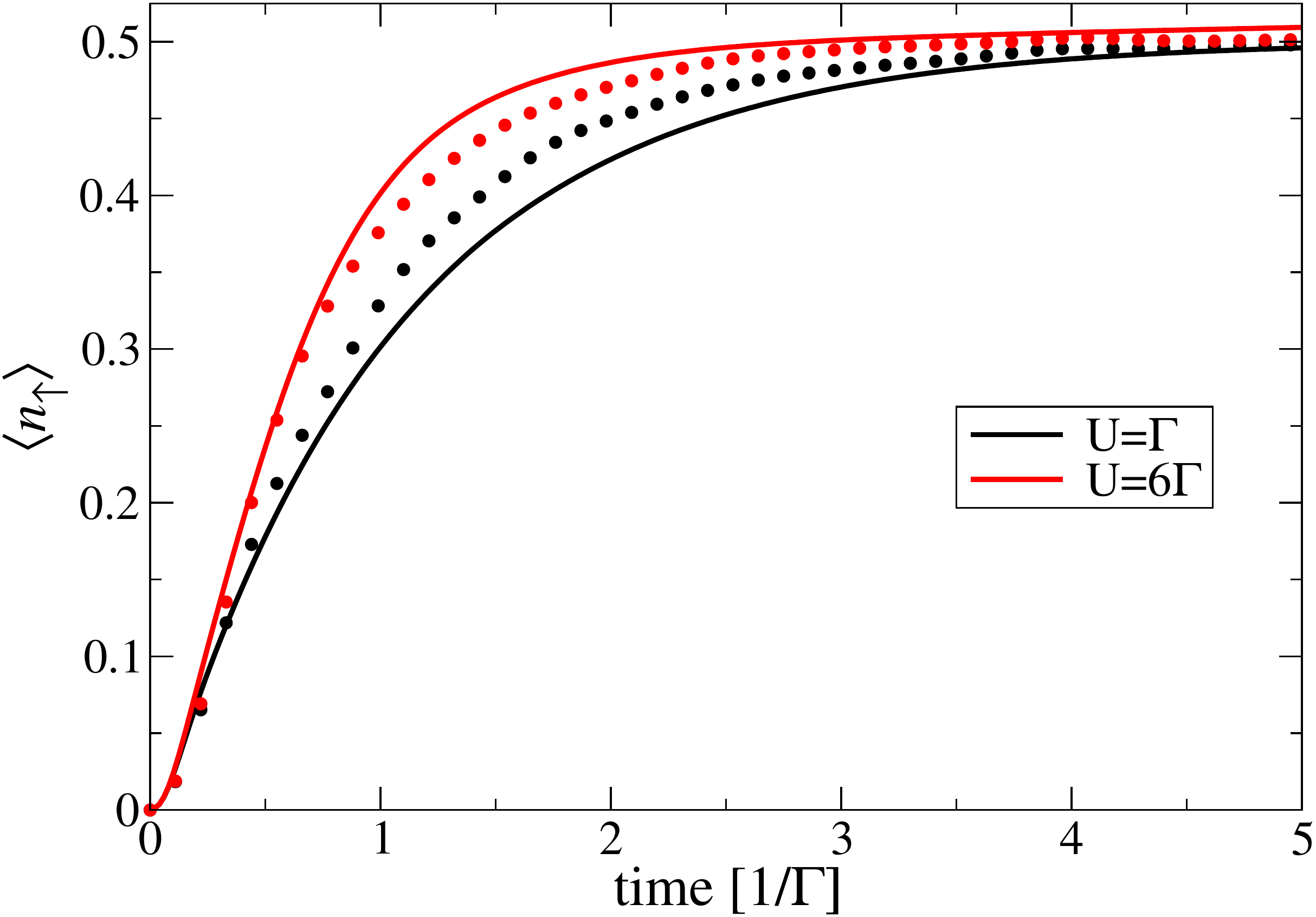}}
\caption{Comparison of the dot population ($\langle
  n_\uparrow+n_\downarrow \rangle$) derivative from the memory-kernel
  formalism~\cite{cohen_memory_2011} (solid lines) and the CQM
  approach (dot symbols) for an initially unoccupied dot for two
  values of the interaction energy ($U=\Gamma$ and
  $U=6\Gamma$). Parameters used: $\Gamma=2\Gamma_{\rm L}=2\Gamma_{\rm
    R}=1$, $T=\Gamma$, $\mu_{\rm L}=-\mu_{\rm R}=\Gamma/2$,
  $\varepsilon_{\uparrow}=\varepsilon_{\downarrow}=-U/2$, $N_{\rm
    tr}=6\times10^4$, and for the CQM results; $\Delta_{\sigma}=0.24$
  for $U=6\Gamma$, and $\Delta_{\sigma}=0.31$ for $U=\Gamma$. }
\label{fig:n_time}
\end{figure}

The agreement between the CQM and the quantum mechanical results are
not limited to steady-state properties. In fact, our mapping also
captures quantitatively the hallmarks of the Coulomb blockade in the
relaxation towards the steady state. Shown in Fig.~\ref{fig:n_time}
is the time dependence of the dot occupation for two different values
of $U$, with a bias voltage of $V_{\rm SD}=\mu_{\rm L}-\mu_{\rm R}=\Gamma$, and
temperature $T=\Gamma$. Here we compare the CQM to numerically
converged real-time stochastically sampled diagrammatic techniques
applied within the reduced density matrix
formalism.\cite{cohen_memory_2011} For both values of $U$ we find that
the full time dependence is in good agreement with the numerically
converged data. In each case, the dot population increases monotonically and the timecales required to reach the long time limit are comparable.

\begin{figure*}[t]
\includegraphics[width=13cm]{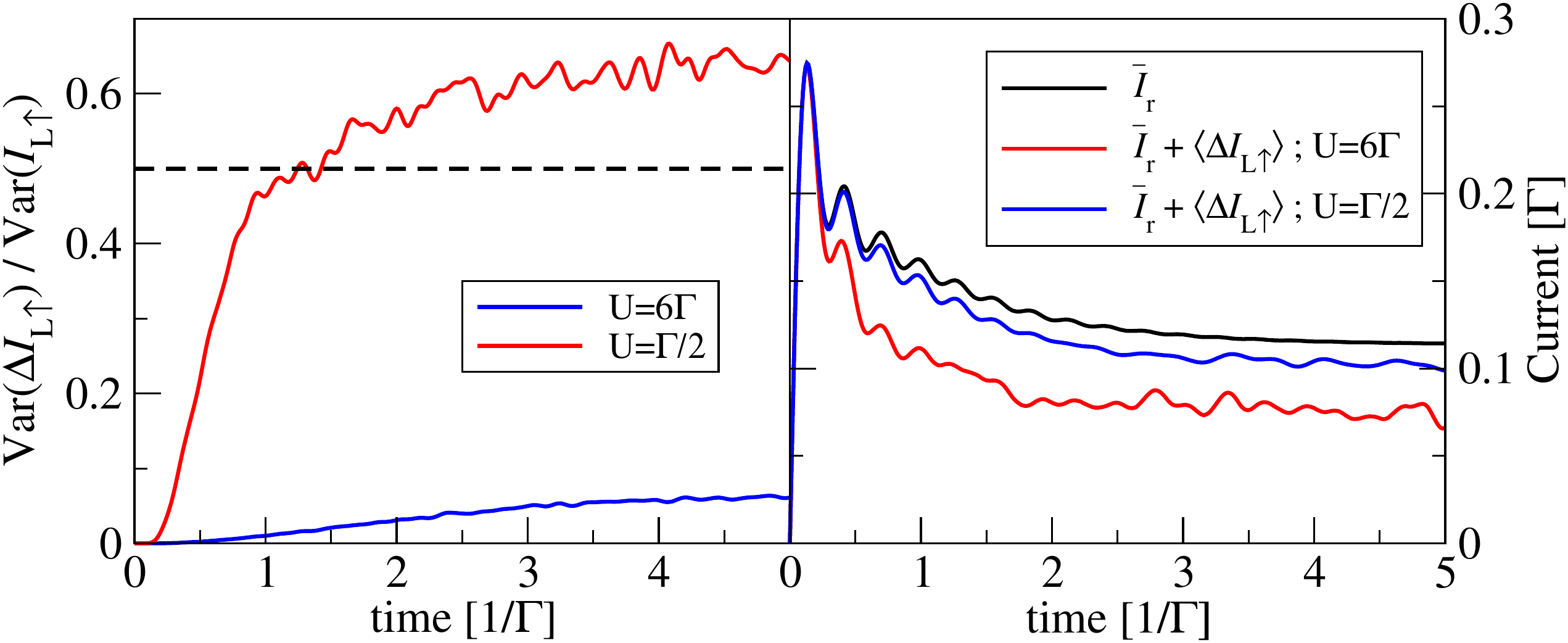}
\caption{Left panel: The ratio of the variances of the left currents
  for spin up as a function of time. The red line is for $U=6\Gamma$
  and the blue for $U=\Gamma/2$. Computational superiority is observed
  below the threshold ratio $1/2$ noted by the black dashed line.
  Right panel: The black line is the exact reference current $(U=0)$,
  the red and blue line are the currents obtained using a reference
  system for $U=6\Gamma$ and $U=\Gamma$, respectively. The parameters are:
  $\Gamma=2\Gamma_{\rm L}=2\Gamma_{\rm R}=1$, $T=0.5\Gamma$,
  $\varepsilon_{\uparrow}=\varepsilon_{\downarrow}=2\Gamma$, $N_{\rm
    tr}=3\times10^4$, and $\mu_{\rm L}=-\mu_{\rm R}=2\Gamma$.  }
\label{fig:var}
\end{figure*}

\section{Reference dynamics for statistical convergence}
\label{sec:ref_system}
When evaluating the classical expected value
Eq.~(\ref{eq:calssical_average}), the integral over the initial
distribution is replaced by averaging over different initial
configurations of the leads that satisfy the Fermi-Dirac distribution.
For a large number of initial conditions, $N_{\rm tr}$, the procedure converges
to the desired distribution and to the exact expectation
value. However, the low dimensional nearly harmonic system generically
requires a large number of initial conditions to statistically
converge the result.  To reduce the number of initial conditions for a
given statistical error, we introduce a reference system whose
expectation value can be determined exactly and inexpensively.
Specifically, the expectation value of an observable $A$ is calculated
according to \beq
\label{eq:avg_ref}
\ave{A}=\ave{A_\mathrm{r} - A_\mathrm{r} + A}=\bar{A}_\mathrm{r} + \ave{\Delta A}
\eeq  
where $A_\mathrm{r}$ is an observable used as the reference, and
$\bar{A}_\mathrm{r}$ is the exact expectation value of $A_\mathrm{r}$
evaluated using a different inexpensive method.  In the limit $N_{\rm tr}
\rightarrow \infty$ we have $\ave{A_\mathrm{r}} \rightarrow
\bar{A}_\mathrm{r}$ and $\ave{A}$ will approach the real expectation
value.  However, for a finite $N_{\rm tr}$, and a smart choice of
$A_\mathrm{r}$, one can reduce significantly the statistical error of
this estimator.  This is clarified by considering the variance using
the reference system

\beq
\text{Var}(\Delta A) =\text{Var}(A)+\text{Var}(A_\mathrm{r})-2\text{Cov}( A,A_\mathrm{r}).
\eeq 
If $A$ and $A_\mathrm{r}$ are correlated, it is possible to have
$\text{Var}(\Delta A) < \text{Var}(A)$.  As we are now propagating
both $A$ and $A_\mathrm{r}$, to reduce the computational effort we
desire that
\beq
\label{eq:var_ratio}
\frac{\text{Var}(\Delta A)}{\text{Var}(A)}<\frac{1}{2}.
\eeq
Given that the sample variance reduces as $1/N_{\rm tr}$, to obtain
computational superiority the ratio in Eq.(\ref{eq:var_ratio}) is
bounded by 1/2. However, improvement in the computational effort can
already be seen for ratios that are above this factor, since typically
propagation of the reference system is not as costly as of the system
of interest.  
Non-interacting or mean-field Hamiltonians that can be solved
analytically serve as examples of reference systems that can reduce
noise for the dynamics of interacting systems.  Other possibilities
include considering dynamics that are generated from some effective
Hamiltonian with the same initial configurations.

Shown in Fig.~\ref{fig:var} are two examples, one in which the
reference system method works good and one in which it fails.  On the
left panel we plot the ratio of the variances of the left current with
spin up as a function of time, and on the right we plot the
corresponding currents.  The reference considered here is the current
calculated for noninteracting systems where an exact solution can be
obtained by direct diagonalization of the single-particle Hamiltonian.
We see that when the reference current, $\bar{I}_{\rm r}$, becomes
very different from the real current, $\bar{I}_{\rm r}+\ave{\Delta
  I}$, (for $U=6\Gamma$ in the figure) the ratio of the variances at
steady-state exceeds its bound $1/2$.  However, when the reference and
the real currents are proximate but still quite different (for
$U=\Gamma/2$ in the figure), we see that the ratio of the variances is
reduced significantly as a consequence of correlations between the
trajectories of the currents.

We note that for parameter regime where the reference and real
currents almost coincide the fluctuations drastically decreased, the
ratio of the variances at steady state reaches as low as $\sim
10^{-4}$.  This implies that for a fixed statistical convergence
threshold, the number of initial conditions decreases by two orders of
magnitude, since each estimate is statistically independent.
One can also note that at short times the reference system always
reduces the fluctuations significantly.  The reason is that we used an
uncorrelated initial condition and thus the short time behavior is set
by $\sim\Gamma^{-1}$.  It takes a certain amount of time for
correlations to build up and for the interacting part in the
Hamiltonian to influence the dynamics, yielding a statistical benefit
for short times even when the steady-state result is far from the
noninteracting limit.

The idea of using a reference system can be extended beyond the
description above. For example, if one wishes to calculate the current
as function of $U$, one can start the evaluation for small $U$ and
increase it ``adiabatically''.  For each calculation of the current,
the previous current (with smaller $U$) can be used as the reference
system. Of course, the exact term in Eq.~(\ref{eq:avg_ref}) is no
longer exact and carry with it some error, but this can still be
beneficial, as trajectories of the different currents are likely to be
correlated given that the change in $U$ is small.

\section{\label{sec:conclusions}conclusions}
We have presented a quasiclassical method to simulate nonequilibrium
dynamics of interacting fermions. We have constructed this map using
the correspondence relation between the commutator and the Poisson
bracket, in order to preserve Heisenberg's equation of motion for
one-body operators.  We have shown that this classical map is complete for
quadratic expectation values under quadratic Hamiltonians and it can
be extended to higher moments accurately.  This feature makes the
study of fluctuations and higher-order correlations accessible.  
\par
For interacting systems, the dynamics is approximated by mapping the
equation of motion and enforcing a quantization rule that determines
for which values of $n_{\sigma}$ the dynamics is influenced by the
Hubbard term.  This, together with a quasiclassical initial
distribution, provides a quantitative agreement with other methods in
regimes where those other methods are known to be accurate.  Thus, a
quantitative description of nonequilibrium currents in the Anderson
model, including their steady-state behavior as illustrated by the
presence of the Coulomb blockade, their fluctuations as encodes in the
second moment of the current, and the relaxation of each to their
steady state can be obtained.

We have also shown a way to enhance the statistical convergence of
this method by introducing a reference system, whose dynamics can be
computed exactly, and averaging the difference between the reference
system and the system of interest.  Provided the reference system is
correlated with the system of interest, fluctuations are reduced
in the averaging procedure, and we have shown that this can increase
the computational efficiency by up to 2 orders of magnitude over naive
sampling.  Together, these results make the quasiclassical method
appealing for studying nonequilibrium phenomena in complex chemical
systems.  Indeed, realistic systems of molecular junctions routinely
operate a low effective temperatures, and finite interaction strengths
rendering other low scaling approximate approaches inaccurate. The
method we have presented here is capable of probing these regimes, at
a small computational cost that scales linearly in the system degrees
of freedom. This should enable studies in correlated transport
behavior in high dimensional, molecular systems, far from equilibrium.

\begin{acknowledgments}
This work was supported by the U.S. Department of Energy, Office of
Basic Energy Sciences, Materials Sciences and Engineering Division,
under Contract No. DEAC02- 05-CH11231 within the Physical Chemistry of
Inorganic Nanostructures Program (KC3103). The authors wish to thank
Lyran Kidon for stimulated discussion.
\end{acknowledgments}

\appendix

\section*{Appendix: Relation to quaternion maps}
\label{ap:A}
The LMM\cite{li2012,li2013} is based on expressing the fermionic
creation and annihilation operators in terms of a set of quaternions:
\beqar
\label{eq:crea/annih}
\aop^{\dagger}=\frac{1}{2}(\sqrt{-1}~\iop -\jop) \\ \nonumber
\aop=\frac{1}{2}(\sqrt{-1}~\iop +\jop).
\eeqar   
The quaternions operators $\iop,\jop$ and $\kop$ satisfy the
anti-commutation relation:
\beqar
\label{eq:quaternions}
\iop \jop &=& -\jop \iop =\kop, \quad \jop \kop=-\kop\jop=\iop, \quad \kop\iop = -\iop\kop =\jop \\ \nonumber
\iop^2 &=& \jop^2=\kop^2=-1.
\eeqar
Using the relations in Eqs.~(\ref{eq:crea/annih}) and
(\ref{eq:quaternions}), quadratic creation and annihilation operators
can be expressed as:
\beqar
\label{eq:quad_crea/anni}
\aop_n^{\dagger}\aop_n &=& \frac{1}{2} +\frac{\sqrt{-1}}{2}\iop_n\jop_n \\ \nonumber
\aop_n^{\dagger}\aop_m &=& \frac{1}{4}\left(-\iopn\iopm-\jopn\jopm + \sqrt{-1}~\left(\iopn\jopm + \iopm\jopn  \right) \right) \\ \nonumber
\aop_n^{\dagger}\aop_m^{\dagger} &=& \frac{1}{4}\left(-\iopn\iopm +\jopn\jopm - \sqrt{-1}~\left(\iopn\jopm - \iopm\jopn  \right) \right) \\ \nonumber
\aop_n\aop_m &=& \frac{1}{4}\left(-\iopn\iopm +\jopn\jopm + \sqrt{-1}~\left(\iopn\jopm - \iopm\jopn  \right) \right) \\ \nonumber
\eeqar 
The commutation relation of two elementary quaternions are then mapped
to a cross product of vectors in phase space,
\beq
\label{eq:map1}
\sqrt{\frac{-1}{2}}~\iop \rightarrow \textbf{r}=\left(\begin{array}{c}
x\\
y
\end{array}\right), \quad \sqrt{\frac{1}{2}}~\jop \rightarrow \textbf{p}=\left(\begin{array}{c}
p_x\\
p_y
\end{array}\right).
\eeq
The CQM replaces the map in Eq.(\ref{eq:map1}) with
\beq
\label{eq:map2}
\sqrt{\frac{\sqrt{-1}}{2}}~\iop \rightarrow u=\left(\begin{array}{c}
x\\
p_x
\end{array}\right), \quad \sqrt{\frac{\sqrt{-1}}{2}}~\jop \rightarrow v=\left(\begin{array}{c}
y\\
p_y
\end{array}\right).
\eeq
This choice implies that
\beqar
\frac{\sqrt{-1}}{2}~\iopn\jopm &\rightarrow & u_n \times v_m =x_m p_{ym}-p_{xn}y_m \\ \nonumber
\frac{\sqrt{-1}}{2}~\iopn\iopm &\rightarrow & u_n \times u_m =x_m p_{xm}-p_{xn}x_m \quad \forall n \neq m \\ \nonumber
\frac{\sqrt{-1}}{2}~\jopn\jopm &\rightarrow & v_n \times v_m =y_m p_{ym}-p_{yn}y_m \quad \forall n \neq m, \\ \nonumber 
\eeqar 
and $\frac{\sqrt{-1}}{2}~\iop\iop = \frac{\sqrt{-1}}{2}~\jop\jop =
-\frac{\sqrt{-1}}{2}$ for $n=m$. The LMM given by Eq.~(\ref{eq:map1})
can be used to map the terms $\aop_n^{\dagger}\aop_n$ and
$\aop_n^{\dagger}\aop_m +\aop_m^{\dagger}\aop_n$. The CQM extend this
to terms like $\aop_n^{\dagger}\aop_m^{\dagger}$, $\aop_n\aop_m$ and
$\aop_n^{\dagger}\aop_m$.


%

\end{document}